\documentclass[10pt]{amsart}
\usepackage{amsmath,amssymb,amsthm}
\usepackage[dvips]{epsfig}

\theoremstyle{plain}
\newtheorem{theorem}{Theorem}[section]

\newtheorem{proposition}[theorem]{Proposition}
\allowdisplaybreaks \numberwithin{equation}{section}

\def\Res{{\rm Res}}

\def\Jac{\mathop{\rm Jac}\nolimits}

\begin{document}

\title[An Extended Abel-Jacobi Map]
{An Extended Abel-Jacobi Map}
\author{H.W. Braden}
\address{School of Mathematics, Edinburgh University, Edinburgh \\
E-mail: hwb@ed.ac.uk}
\author{Yu. N. Fedorov}
\address{Department de Mathematica I\\
Universitat Politecnica de Catalunya, Spain \\
E-mail: {Yuri.Fedorov@upc.edu}}
\begin{abstract}
We solve the problem of inversion of an extended Abel-Jacobi map
$$
\int_{P_{0}}^{P_{1}}\omega +\cdots+\int_{P_{0}}^{P_{g+n-1}}\omega
={\bf z} , \qquad \int_{P_{0}}^{P_{1}}\Omega_{j1}+\cdots
+\int_{P_{0}}^{P_{g+n-1}}\Omega _{j1} =Z_{j},\quad j=2,\dots,n ,
$$
where $\Omega_{j1}$ are (normalised) abelian differentials of the
third kind. In contrast to the extensions already studied, this
one contains meromorphic differentials having a common pole $Q_1$.
This inversion problem arises in algebraic geometric description
of monopoles, as well as in the linearization of integrable
systems on finite-dimensional unreduced coadjoint orbits on loop
algebras.
\end{abstract}

\maketitle

\section{Introduction. }
The purpose of this note is to present and invert an extension of
the classical Abel-Jacobi theorem that  was recently encountered
in the construction of nonabelian magnetic monopoles \cite{be06}
and also appeared in the linearization of integrable systems on
finite-dimensional unreduced coadjoint orbits on various loop
algebras (\cite{AHH_CMP, Gavr, Gag, RST}). We believe this
extension is of independent interest and will now describe it,
placing it in context.

Let $\Gamma$ be a Riemann surface of genus $g$ with a canonical
homology basis $\{a_i,b_i\}_{i=1}\sp{g}$ and holomorphic
differentials $\omega =(\omega _{1},\dots,\omega _{g})$ normalized
so that $\oint_{a_i}\omega_j=\delta_{ij}$. (We follow the
conventions of \cite{fk80}.)  Let $\Lambda$ be the rank $2g$
lattice in $ {\mathbb C}^g =(z_1,\dots,z_g)={\bf z}$ generated by
the vectors of periods of $\omega$ with respect to $a_1,\dots,
b_g$ and $J(\Gamma)={\mathbb C}^g/\Lambda$ be the Jacobian of the
curve $\Gamma$. The Abel map $A(P)=\int_{P_0}\sp{P}\omega$ gives a
well defined holomorphic map of $\Gamma$ into $J(\Gamma)$ that may
be extended to a map of divisors. Abel's theorem identifies the
divisor of a meromorphic function on $\Gamma$ with degree zero
divisors in the kernel of $A$ and Jacobi's (inversion) theorem
says that every point on the Jacobian $J(\Gamma)$ is expressible
as the image under the Abel map of a positive divisor of degree
$g$: that is, for all ${\bf z}\in J(\Gamma)$ we may find
$P_1,\dots,P_g\in \Gamma$ such that
$$
\int_{P_{0}}^{P_{1}}\omega +\cdots+\int_{P_{0}}^{P_{g}}\omega  ={\bf z}.
$$
Analytically this inversion makes use of Riemann's theta
function. The Abel-Jacobi theorems play a central role in the theory
of Riemann surfaces and find important application in the solution
of integrable systems. The modern approach to integrability places
special attention on algebraically completely integrable systems:
systems whose real invariant tori may be extended to complex
algebraic tori (Abelian varieties) related to an algebraic curve and
whose complex phase flows become straight line motion on the tori. A
Lax equation, for example, will frequently result in a curve
$\Gamma$, the relevant Abelian variety is the Jacobian $J(\Gamma)$
and the dynamics is linearizable on the Jacobian. In this setting
the physical solution is obtained by solving the Jacobi inversion problem.

Although many known integrable systems fit into the framework just
described, some extensions have been required over the years. In
particular, if one wishes to describe the rotation matrices of
some integrable tops (\cite{BBEIM, Gav_Z, fed99}) or to obtain
asymptotic (heteroclinic) solutions (such us umbilic geodesic
trajectories on a triaxial ellipsoid) (\cite{Er, fed99}), one
arrives at quadratures that, in addition to the holomorphic
differentials already noted, involve various meromorphic
differentials of the third kind, $\Omega_{P_+,P_-}$, with simple
poles at $P_\pm$ and having residues $\pm1$. Clebsch and Gordan
\cite{cg86} described an appropriate extension of Abel-Jacobi
theory to take these into account. Namely, suppose
$X_1,Y_1,\dots,X_s,Y_s$ are distinct pairs of points on $\Gamma$
and $\Omega_{X_i,Y_i}$ the corresponding differentials of the
third kind. We will assume throughout that such differentials are
normalized to have vanishing $a$ periods. Then (for
$P_0\not\in\{X_1,Y_1,\dots,X_s,Y_s\}$) the extended Abel map
$(\int_{P_0}\sp{P}\omega,\int_{P_0}\sp{P}\Omega_{X_1,Y_1},\dots,
\int_{P_0}\sp{P}\Omega_{X_s,Y_s})$ admits the inversion of
\begin{equation} \label{ext_aj}
\int_{P_{0}}^{P_{1}}\omega +\cdots+\int_{P_{0}}^{P_{g+s}}\omega
={\bf z} , \qquad \int_{P_{0}}^{P_{1}} \Omega_{X_i,Y_i} +\cdots
+\int_{P_{0}}^{P_{g+s}}\Omega _{X_i,Y_i} =Z_{i},\quad i=1,\dots s.
\end{equation}
In this setting the Jacobian of $\Gamma$ is replaced by a
noncompact Abelian variety, an algebraic group, and generalized
theta functions replace the Riemann theta function in the analytic
inversion. This result may be viewed as a limit of the classical
Abel-Jacobi theory for a genus $g+s$ curve as $s$ $a$-cycles pinch
to zero resulting in a singular surface. ( The $s=1$ account of
this may be found in \cite{fa73}, for a more general case one can
consult \cite{fed99,Al_Fed_Inverse}.)
\medskip

In this paper we consider a special degeneration of the map
(\ref{ext_aj}), when all the meromorphic differentials have a {\it
common} pole. Namely, let $Q_1,\dots,Q_n$ be distinct points of
our surface $\Gamma$ and let $\Omega _{21},\dots,\Omega_{n1}$ be
normalized meromorphic differentials of the third kind (with
vanishing $a$ periods) having pairs of simple poles at points
$\left( Q_{2},Q_{1}\right), \dots, \left( Q_{n},Q_{1}\right)$
respectively, and such that
$$
 \Res_{Q_2}\,\Omega_{21}=\cdots = \Res_{Q_n}
\,\Omega_{n1}=1 \quad \mbox{and} \quad \Res_{Q_1}
\,\Omega_{21}=\cdots = \Res_{Q_1} \Omega_{n1}=-1.
$$
Let, as above, ${\bf z}=(z_{1},\dots,z_{g})\in\mathbb{C}\sp{g}$
and $ {\bf \hat z} =({\bf z},Z_2,\dots,
Z_n)\in\mathbb{C}\sp{g+n-1}$. (Note that now there is no
$Z_1$-variable here.) Then (for $P_0\not\in\{Q_1,\dots,Q_n\}$) the
extended Abel map ${\mathcal A}
(P)=(\int_{P_0}\sp{P}\omega,\int_{P_0}\sp{P}\Omega_{21},\dots,
\int_{P_0}\sp{P}\Omega_{n1})$ admits the inversion of
\begin{equation}
\int_{P_{0}}^{P_{1}}\omega +\cdots+\int_{P_{0}}^{P_{g+n-1}}\omega
={\bf z} , \qquad \int_{P_{0}}^{P_{1}}\Omega_{j1}+\cdots
+\int_{P_{0}}^{P_{g+n-1}}\Omega _{j1} =Z_{j},\quad j=2,\dots,n
\label{genaj}.
\end{equation}
In our extension there is obvious overlap with the Clebsch-Gordan theory for $s=1$ and
$n=2$. Note that the theta-functional formulas describing the
inversion of (\ref{ext_aj}) for $s>1$ do not survive under the
limit $Y_1,\dots, Y_s \to Q_1$; the pinchings needed to view this
result as a degeneration of the usual Abel-Jacobi theory are
delicate (see also Remark 1 below).
Apparently, the inversion problem in this case was not treated in the literature, and
we will give a direct solution to this problem rather than consider such a limit.

A precise formulation of our result and the generalized theta
functions which arise in the analytical inversion will be given in
the next section. Here we remark that the extension (\ref{genaj})
naturally appears in the linearization of integrable systems on
finite-dimensional unreduced coadjoint orbits of various loop
algebras (see, among others, \cite{AHH_CMP, Gavr, RST}). The
systems are described by $n\times n$ matrix Lax pairs with a
rational parameter, and the spectral curve $\Gamma$ is an $n$-fold
cover of ${\mathbb P}\sp1$ with infinite points $\infty_1, \dots,
\infty_n$. The corresponding quadratures involve a complete set of
$g$ holomorphic differentials on $\Gamma$, as well as
differentials with pairs of simple poles
$(\infty_1,\infty_2),\dots, (\infty_1,\infty_n)$. For a generic
choice of Darboux coordinates on the unreduced orbits, the
quadratures describe evolution of a divisor of $g+n-1$ finite
points on $\Gamma$ and thereby have the form
(\ref{genaj})\footnote{Note that a special choice of such
coordinates fixes $n-1$ points of the divisor at infinity, and the
the first equation in (\ref{genaj}) becomes a standard Abel map
admitting the inversion by the Riemann theta-function. This
approach was pursued in \cite{AHH_CMP}.}

The same extension appears in algebraic completely integrable
system associated to the BPS limit of $su(2)$ magnetic monopoles
and $\Gamma$ is the attendant curve, which has a natural spatial
interpretation \cite{hitchin83}. In common with many such
integrable systems, Baker-Akhiezer functions on the curve play a
prominent role. Such functions are a slight extension to the class
of meromorphic functions that allow essential singularities at a
finite number $n\ge1$ of points; they have many properties similar
to those of meromorphic functions. While for a meromorphic
function one needs to prescribe $g+1$ poles in the generic
situation, a non-trivial Baker-Akhieser function exists with $g$
arbitrarily prescribed poles on a surface of genus $g$. The
construction of such functions may be made rather explicitly in
terms of Riemann's theta function and involves an otherwise
unspecified generic divisor of degree $g+n-1$. In the monopole
setting, a solution to the inversion problem (\ref{genaj}) means
that this unspecified divisor will have no physical consequences:
such should be the case, and consequently led to the present
theory developed in \cite[\S3.2]{be06}.

\section{Notation and Results}

Let $\hat \Gamma$ be the $4g$-sided polygon with symbol $a_1 b_1
a_1\sp{-1}b_1\sp{-1}\dots a_g b_g a_g\sp{-1}b_g\sp{-1}$ obtained by
dissecting $\Gamma$ along the cycles of the chosen homology basis.
We assume that our points $Q_i$ do not lie on any of our canonical
cycles. Let $\gamma_i$ be (disjoint) cycles around the points $Q_i$
lying in $\hat \Gamma$ oriented so that
$\oint_{\gamma_i}\Omega_{i1}=\Res_{Q_i}\Omega_{i1}=1$ and
$\oint_{\gamma_1}\Omega_{i1}=\Res_{Q_1}\Omega_{i1}=-1$ (for $2\le
i\le n$). Choose a point of $\partial\hat \Gamma$ as a vertex and
let $\hat\Gamma_c$ denote the domain $\hat \Gamma$ with (disjoint)
cuts from this vertex to each of the points $Q_i$. Then
$\int_{P_0}\sp{P}\Omega_{j1}$ is single valued on $\hat \Gamma_c$.

The generalized map ${\mathcal A} (P)$ has $2g+n-1$ independent
period vectors corresponding to the $2g$ canonical cycles
$a_i,b_i$ and cycles $\gamma_2,\dots,\gamma_n$. The factor of
$\mathbb{C}^{g+n-1}$ by the lattice generated by these vectors
will be called the generalized Jacobian Jac$(\Gamma; Q_1,\dots,Q_n)$.

Introduce (for $n\ge2$) the {\it generalized theta function}
\begin{align}
 \varTheta_n({\bf \hat z}) & \equiv \varTheta({\bf z},Z_{2},\dots,Z_{n}) \nonumber \\
& = \sum_{i=2}^n \exp (Z_i-\mathcal{K}_i- \Delta_i ) \, \theta
\left( {\bf z}-K-\mathcal{S}+ A(Q_i)\right )
- \theta \left({\bf z}  -K-\mathcal{ S}+ A(Q_1) \right ) ,  \label{VarTheta}  \\
\Delta_i & = \sum_{k\ne i, \, k\ne 1} \int_{P_0}^{Q_k} \Omega_{i1},
\quad \mathcal{ S}= \sum_{i=1}^n A(Q_i)\, , \label{pars}
\end{align}
where, as above, $A(P)=\int_{P_0}^P (\omega_1,\dots, \omega_g)$ is
the customary Abel map and $\theta({\bf z})$ is the canonical
Riemann theta-function with the Riemann period matrix $\tau$ of $\Gamma$,
\begin{equation}
\theta(\boldsymbol{z})\equiv\theta(\boldsymbol{z};\tau)
=\sum_{\boldsymbol{n}\in \mathbb{Z}^g} \exp(\imath\pi
\boldsymbol{n}^T \tau\boldsymbol{n}+2 \imath\pi \boldsymbol{z}^T
\boldsymbol{n}) .
\end{equation}
We will suppress throughout the dependence on $\tau$.
The vector $K$ in (\ref{VarTheta}) is
the vector of Riemann constants (see e.g., \cite{fa73}), while $\mathcal{K}_i$ ($i\ge2$)
are the following constants (fixed in our discussion below)
\begin{equation}
\mathcal{K}_i=\sum_{k=1}\sp{g}\oint_{a_k}\omega_k(P)\int_{P_0}\sp{P}\Omega_{i1}+
\oint_{b_k}\Omega_{i1}. \label{defki}
\end{equation}

Observe that for $n\ge3$ there is a recursive description
\begin{equation}
\varTheta_n({\bf \hat z})=\exp (Z_n-\mathcal{K}_n- \Delta_n ) \,
\theta \left( {\bf z}-K-\sum_{i=1}\sp{n-1}A(Q_i)\right
)+\varTheta_{n-1}({\bf \hat z}-\mathcal{A}(Q_n)). \label{inddef}
\end{equation}

We will establish two inversion theorems based on the function
$$
f(P)= \varTheta_n ( {\bf \hat z}- \mathcal{ A} (P)).
$$
We first observe

\begin{proposition} \label{propquasi}
\begin{itemize}
\item[1).] Although $f(P)$ is not single-valued on the curve
$\Gamma$, its zeros do not depend on the choice of path in the
Abel map $\mathcal{A}(P)$.

\item[2).] The differential $d\ln f(P)$ is regular at $Q_1$ and
has poles of residue $-1$ at each of $Q_2,\dots,Q_n$.

\item[3).] The function $f(P)$ has precisely $g+n-1$ zeros
(possibly with multiplicity).
\end{itemize}
\end{proposition}
With this our generalized inversion theorems are then

\begin{theorem} Let ${\bf \hat z}=({\bf z},Z_2,\dots, Z_n)$ be such that the function
$ f(P)= \varTheta_n ( {\bf \hat z}- \mathcal{ A} (P)) $ does not
vanish identically on  $\hat\Gamma \setminus\{Q_1,\dots,Q_n\}$. In
this case, if ${\bf \hat z}$ is the right hand side of
(\ref{genaj}), the $g+n-1$ zeros give a unique solution to the
inversion of the extended map (\ref{genaj}). \label{genzeros}
\end{theorem}

\begin{theorem} \label{gendiv}
Let ${\bf \hat z} \in \{ \varTheta_n ( {\bf \hat z} )=0\}$ be such that
$f(P)=\varTheta_n ( {\bf \hat z} - \mathcal{A} (P))$ does not
vanish identically on $\hat\Gamma \setminus\{Q_1,\dots,Q_n\}$.
Then there exists a unique positive divisor $P_1+\cdots+
P_{g+n-2}$ such that
\begin{equation} \label{theta0}
{\bf \hat z} =\mathcal{ A} (P_1) +\cdots+ \mathcal{A} (P_{g+n-2}).
\end{equation}
\end{theorem}

\paragraph{Remark 1.} According to \cite{cg86, fed99}, the inversion
of the generalized map (\ref{ext_aj}), that includes meromorphic
differentials $\Omega_{X_i,Y_i}$ with {\it different} poles, is
given by zeros of the generalized theta-function
$$
\tilde{\theta} ( \tilde {\bf z}- \mathcal{A}(P) -\tilde K), \qquad
\tilde {\bf z}=( {\bf z}, Z_1,\dots,Z_s)^T, \quad \tilde {K}=(K,
k_1,\dots,k_s)^T, \quad k_j=\mbox{const} ,
$$
which is represented by the sum of $2^{s}$ terms,
\begin{gather}
\tilde{\theta} ( {\bf z},Z) =\sum_{{\varepsilon}_{1},\dots{\varepsilon}_{s}=\pm 1}
\exp\left(\frac12 ({\varepsilon},Z)+\frac14 ({\varepsilon},{\bf S}{\varepsilon})\right)
\theta\left( {\bf z} + \frac 12 \varepsilon_1 q_1+\cdots+
\frac 12 \varepsilon_s q_s\right ) , \label{2s}
\\
{\varepsilon}=({\varepsilon}_{1},\ldots,{\varepsilon}_{s})^{T},\quad
({\varepsilon},Z)=\sum^{s}_{l=1}{\varepsilon}_{l}Z_{l}, \quad
({\varepsilon},{\bf S}{\varepsilon})=\sum^{s}_{l,r=1, l\ne r}
S_{lr}{\varepsilon}_{l}{\varepsilon}_{r}.  \nonumber
\end{gather}
where $q_l=(q_{l1},\dots,q_{lg})^T$ is the $b$-period vector of
$\Omega_{X_{l}Y_{l}}$ and $S_{lr}$ are given by
\begin{gather}
q_{li}=\int^{Y_l}_{X_l}\bar{\omega}_{i}=\oint_{b_{i}} \Omega_{X_{l}Y_{l}} ,\quad
S_{lr} =\int^{Y_{l}}_{X_{l}}\Omega_{X_{r}Y_{r}}
=\int^{Y_{r}}_{X_{r}}\Omega_{X_{l}Y_{l}}=S_{rl}\; (l\neq r), \label{Sij} \\
i=1,\dots,g,\quad l,r=1,\dots,s . \nonumber
\end{gather}
As follows from (\ref{Sij}), in the limit $Y_1,\dots Y_s \to Q_1$,
which transforms the extended map (\ref{ext_aj}) to (\ref{genaj})
with $n=s+1$, the integrals $S_{lr}$ become infinite and the theta
function $\tilde{\theta} ( {\bf z},Z)$ becomes singular.
Alternatively, one may try to divide the sum (\ref{2s}) by certain
exponents of $S_{lr}$ {\it before} taking the limit to obtain a
finite expression instead.  However, this procedure involves
several delicate steps, which appeared to be more complicated than
the direct approach proposed in this paper.
\medskip

Our proof is inductive. In section 3 we will establish Theorem
\ref{genzeros} for the case $n=2$. Section 4 contains the
inductive step and completes the proof of Theorem \ref{genzeros}.
Here we also establish the formulae (\ref{defki}) and give a proof
of Theorem \ref{gendiv} closing with some remarks on the
generalized theta divisor. We conclude the section by proving
Proposition \ref{propquasi}.

\begin{proof}[Proof of Proposition \ref{propquasi}]

1). Our choice of normalization of the differentials $\omega_i$
and $\Omega_{i1}$ together with the invariance $\theta({\bf
z}+{\bf e}_k)=\theta({\bf z})$, where ${\bf e}_k$ is the vector
with $1$ in the $k$-th entry and zero otherwise, says that $f(P)$
is invariant as $P$ encircles an $a$-cycle. Similarly encircling a
$\gamma$-cycle at most changes the exponential factors by $2\pi
i$'s and we again have invariance in this case. Finally consider
encircling a $b$-cycle, say $b_k$. The quasi-periodicity of the
Riemann theta function
$$
\theta \left( {\bf z}+\oint_{b_k}\omega;\,\tau \right)
=\exp[-i\pi(\tau_{kk}+2 z_k)]\,\theta({\bf z};\,\tau)$$ means that
the $j$-th term of the sum on the right-hand side of
(\ref{VarTheta}) changes by the phase
$$2\pi i\left({\bf z}-K-A(P)-\mathcal{S}+A(Q_j)\right)_k-\oint_{b_k}\Omega_{j1}
-i\pi\tau_{kk}.
$$
Now the bilinear relation
\begin{equation}\label{thirdkn}2\pi i(
A(Q_j)-A(Q_1))=2\pi i
\int_{Q_1}\sp{Q_j}\omega_k=\oint_{b_k}\Omega_{j1} \end{equation}
means this may be rewritten as \begin{equation}\label{phasen}2\pi
i\left({\bf z}-K-A(P)-\mathcal{S}+A(Q_1)\right)_k
-i\pi\tau_{kk}\end{equation} and this is just the change of phase
of the remaining term on the right-hand side of (\ref{VarTheta}).

Therefore $\varTheta_n({\bf \hat z}-\mathcal{A}(P))$ changes by
the phase (\ref{phasen}) and hence is a quasi-periodic function.
The quasi-periodicity of $f(P)$ means that although the function
is not single-valued on the curve $\Gamma$ its zeros do not depend
on the choice of path in the Abel map $\mathcal{A}(P)$.
\medskip

2). Next let us consider the behaviour of $f(P)$ as $P\sim Q_j$
($j=1,\dots,n$) focusing initially on the case $j\ge2$. The $i$-th
term in the sum of on the right-hand side of (\ref{VarTheta}) for
$f(P)$ contains the exponential term
$\exp(-\int_{P_0}\sp{P}\Omega_{i1})$. This is regular as
$P\rightarrow Q_j$ for $j\ne i$ and so as $P\sim Q_j$
\begin{equation*}d\ln\varTheta_n({\bf \hat z}-\mathcal{A}(P))\sim d
\ln\left(\alpha \exp(-\int_{P_0}\sp{P}\Omega_{j1})+\beta\right).
\end{equation*}
Generically the coefficient $\alpha$ is non-vanishing. Now as
$P\sim Q_j$ (with local coordinates $\tau$)
\begin{equation*}
 \exp \bigg(-\int_{P_0}\sp{P}\Omega_{j1}\bigg )\sim
\exp\left(-\int\sp{\tau}\frac{dz}{z} \right) \sim\frac1{\tau}
\end{equation*}
and consequently
\begin{equation}
d\ln\varTheta_n({\bf \hat z}-\mathcal{A}(P))\sim
d\ln\left(\frac{\alpha \tau+\beta}{\tau}\right)
\label{cnresb}
\end{equation}
has residue $-1$ at $Q_j$ ($j\ge2$). Finally, as $P\sim Q_1$ (with
local coordinate $\tau$) each of the exponential terms
\begin{equation} \label{Q1_limit}
 \exp\left( -\int_{P_0}\sp{P}\Omega_{j1}\right) \sim
\exp\left( \int\sp{\tau}\frac{dz}{z} \right) \sim \tau
\end{equation}
give a regular vanishing and so as $P\sim Q_{1}$
\begin{equation}
d\ln\varTheta_n({\bf \hat z}-\mathcal{A}(P))\sim d\ln (\alpha
\tau+\beta)\label{c2resa}.
\end{equation}
Generically $\beta\ne0$
and this is regular at $Q_1$  and hence has vanishing residue.
Indeed, considering now all of (\ref{VarTheta}), we find that
\begin{equation}
\lim_{P\rightarrow Q_1}\varTheta_n({\bf
\hat z}-\mathcal{A}(P))= - \theta \left({\bf z}
-K-\sum_{i=1}\sp{n} A(Q_i) \right ) \equiv\beta.
\label{cnresc}
\end{equation}
Thus we have established that $d\ln \varTheta({\bf \hat
z}-\mathcal{A}(P))$ is regular at $Q_1$ and has poles of residue
$-1$ at each of $Q_2,\dots,Q_n$.
\medskip

3). To count the number of zero's we evaluate
$$
\frac{1}{2\pi i}\oint_{\partial\hat\Gamma_c}
 d\ln\varTheta_n({\bf \hat z}-\mathcal{A}(P))=
 \frac{1}{2\pi  i}\left[\sum_{k=1}\sp{g}\oint_{a_k+a_k\sp{-1}+b_k+b_k\sp{-1}}-\sum_{r=1}\sp{n}\int_{\gamma_r}
 \right] d\ln\varTheta_n({\bf \hat z}-\mathcal{A}(P))
$$ (upon noting that we encircle the $Q_i$'s in the opposite direction to the outer
boundary). Now utilizing the residues just determined the second
term in this expression contributes
$$
-\frac{1}{2\pi i}\sum_{r=1}\sp{n}\int_{\gamma_r}
d\ln\varTheta_n({\bf \hat z}-\mathcal{A}(P)) =
-\sum_{P\in\{Q_1,\dots,Q_n\}}\Res_P\, d\ln\varTheta_n({\bf \hat z}-\mathcal{A}(P)  =n-1.
$$
The first term on the other hand is simplified using the
quasi-periodicity properties of part (1) of the proposition.
Letting $f\sp{-}$ denote the value of $f$ on the cycles
$a_k\sp{-1}$ or $b_k\sp{-1}$ we have that if $P$ lies on $a_k$
then from (\ref{phasen})
$$
d\ln f\sp-(P)=d\ln f(P)-2\pi i\,\omega_k(P)
$$
while if $P$ is on $b_k$ then
$$
f\sp-(P)=f(P).
$$
Thus
\begin{align*}
\frac{1}{2\pi i}\oint_{a_k+a_k\sp{-1}}\,d\ln f(P) &= \frac{1}{2\pi
 i}\oint_{a_k}\left[d\ln f(P)
-d\ln f\sp-(P)\right] = \oint_{a_k}\omega_k(P)=1
\end{align*}
and
$$
\frac{1}{2\pi i}\oint_{b_k+b_k\sp{-1}}\,d\ln f(P)=0.$$
Together these yield
$$
g=\frac{1}{2\pi
 i}\sum_{k=1}\sp{g}\oint_{a_k+a_k\sp{-1}+b_k+b_k\sp{-1}}\,
d\ln\varTheta_n({\bf \hat z}-\mathcal{A}(P)  $$
and so the number of zeros of $f(P)$ is then $g+n-1$.
\end{proof}

\section{Clebsch's Case: $n=2$}
We now make the first step in the proof of Theorem 2.2. Namely,
following \cite{cg86}, consider the extended Abel map ${\mathcal
A} \left( P_{1},\dots,P_{g+1}\right) \mapsto {\mathbb C}^{g+1}={\bf \hat z}=({\bf z}, Z_{2})$,  defined by
\begin{eqnarray}
\int_{P_{0}}^{P_{1}}\omega +\cdots+\int_{P_{0}}^{P_{g+1}}\omega  &=&{\bf z} ,\label{cleb2a} \\
\int_{P_{0}}^{P_{1}}\Omega_{21}+\cdots +\int_{P_{0}}^{P_{g+1}}\Omega_{21} &=&Z_{2}  ,\label{cleb2b}
\end{eqnarray}
where $P_0\ne Q_{1,2}$. We are going to show that its inversion is given in terms of the generalized
theta function that comes from (\ref{VarTheta}), (\ref{pars}),
\begin{equation}\begin{split}
 \varTheta_2({\bf \hat z}-\mathcal{A}(P))
& =  \exp \left( Z_2- \mathcal{K}_2-\int_{P_0}\sp{P}\Omega_{21} \right ) \,
\theta \left( {\bf z}-K-A(P)- A(Q_1)\right )\\&\qquad - \theta\left({\bf z} -K-A(P)- A(Q_2) \right ) ,
\end{split}\label{VarTheta2}
\end{equation}
where
\begin{equation} \label{K2}
\mathcal{K}_2=\sum_{k=1}\sp{g}\oint_{a_k}\omega_k(P)\int_{P_0}\sp{P}\Omega_{21}+
\oint_{b_k}\Omega_{21}.
\end{equation}
That is, we now establish the $n=2$ case of Theorem \ref{genzeros}.

\begin{proposition} \label{inv_clebsch}
Let ${\bf \hat z}$ be the right hand side of the extended map (\ref{cleb2a}), (\ref{cleb2b})
 such that $f(P)=\varTheta_2 ( {\bf \hat z}- \mathcal{ A} (P))$
does not vanish identically on  $\hat\Gamma\setminus\{Q_1,Q_2\}$.
Then $f(P)$ has precisely $g+1$ zeros (possibly with multiplicity) giving a unique
solution to the inversion of the map.
\end{proposition}

\noindent{\it Proof.} The number of zeros and quasi-periodicity
of $f(P)$ have already been described in Proposition
\ref{propquasi} and it remains to show that this function uniquely
solves the inversion.

First let us suppose $P_{g+1}=P_0$. Then (\ref{cleb2a}) becomes
the usual Abel-Jacobi map and $\bf z$ determines the divisor
$P_1+\dots+P_g$ uniquely. In this case $Z_2$ becomes a function of
${\bf z}$. Following \cite{cg86}, we express (\ref{cleb2b}) as a
sum of residues and obtain
\begin{align}
Z_2 & =\int_{P_{0}}^{P_{1}}\Omega_{21}+\cdots
+\int_{P_{0}}^{P_{g}}\Omega_{21} =\sum_{k=1}\sp{g}\Res_{P=P_k}\left(\int_{P_{0}}^{P}\Omega_{21}\right)
d\ln\theta({\bf z}-K-A(P))  \nonumber  \\
&=\frac{1}{2\pi i}\oint_{\partial\hat\Gamma_c}
\left(\int_{P_{0}}^{P}\Omega_{21}\right) d\ln\theta({\bf z}-K-A(P))  \nonumber  \\
&=\frac{1}{2\pi i}\oint_{\partial\hat\Gamma}
\left(\int_{P_{0}}^{P}\Omega_{21}\right) d\ln\theta({\bf z}-K-A(P))-
\Res_{P\in\{Q_1,Q_2\}} \left(\int_{P_{0}}^{P}\Omega_{21}\right)
d\ln\theta({\bf z}-K-A(P)) \nonumber  \\
&=\frac{1}{2\pi i}\oint_{\partial\hat\Gamma}
\left(\int_{P_{0}}^{P}\Omega_{21}\right) d\ln\theta({\bf z}-K-A(P))+
\Res_{P\in\{Q_1,Q_2\}}\, \Omega_{21}(P) \,\ln\theta({\bf z}-K-A(P)) \nonumber  \\
&= \hat{\mathcal K}_2 +\ln \frac{\theta({\bf z}-K-A(Q_2) )}{\theta({\bf z}-K-A(Q_1) ) }, \label{trans_Z}
\end{align}
where we set
\begin{equation} \label{pre_K2}
\hat{\mathcal K}_2 =\frac{1}{2\pi i}\oint_{\partial\hat\Gamma}
\left(\int_{P_{0}}^{P}\Omega_{21}\right) d\ln\theta({\bf
z}-K-A(P)) = \frac{1}{2\pi i}\sum_{k=1}\sp{g}\oint_{a_k+a_k\sp{-1}+b_k+b_k\sp{-1}}\,h(P)dg(P).
\end{equation}
Here
$$
h(P) = \int_{P_0}\sp{P}\Omega_{21} , \quad g(P)= \ln\theta({\bf z}-K-A(P))
$$
and in  the derivation of (\ref{trans_Z}) we have used the single
valuedness of $h(P)$ and $ g(P)$ on $\hat\Gamma_c$.

Now let us consider in (\ref{cleb2a}), (\ref{cleb2b}) the generic
positive divisor $P_1+\cdots +P_{g+1}$ where $P_j\ne Q_1,Q_2$.
This situation can be reduced to the previous one by replacing
${\bf \hat z}$  by ${\bf \hat z}-\mathcal{A}(P_j)$. Then the
relation (\ref{trans_Z}) gives rise to
$$
\exp\left(Z_2-\hat{\mathcal K}_2- \int_{P_0}^{P_j}\Omega_{21}\right)
= \frac{\theta({\bf z}-K-A(Q_2)-A(P_j) )}{\theta({\bf z}-K-A(Q_1)-A(P_j) ) },
\qquad j=1,\dots,g+1 .
$$
That is, if $\hat{\mathcal K}_2=\mathcal{K}_2$, the function
$f(P)=\varTheta_2 ( {\bf \hat z}- \mathcal{ A} (P))$
vanishes at $P_1,\dots, P_{g+1}$ with the same multiplicity.
According to Proposition \ref{propquasi}, in our case $f(P)$ has
precisely $g+1$ zeros, which therefore must coincide with
$P_1,\dots, P_{g+1}$. This proves Proposition \ref{inv_clebsch}.

It remains only to show that $\hat{\mathcal K}_2$ coincides with
$\mathcal{K}_2$ given by expression (\ref{K2}). Let $h\sp{-}$,
$g\sp{-}$  denote the value of $h$, respectively $g$ on the cycles
$a_k\sp{-1}$ or $b_k\sp{-1}$. Then, if $P$ lies on $a_k$,
$$
h\sp-(P)=h(P)+\oint_{b_k}\Omega_{21},\qquad g\sp-(P)=g(P)-2\pi i\,\omega_k(P)\,
$$
while if $P$ is on $b_k$ then
$$
h\sp-(P)=h(P),\qquad g\sp-(P)=g(P).
$$
Thus
\begin{align*}
\oint_{a_k+a_k\sp{-1}}\,h(P)dg(P) &= \oint_{a_k}\left[h(P)dg(P)
-h\sp-(P)d g\sp-(P)\right]\\
&=2\pi i\oint_{a_k}\omega_k(P)\int_{P_0}\sp{P}\Omega_{21}-\oint_{b_k}\Omega_{21}
\oint_{a_k}dg(P)+2\pi i\oint_{b_k}\Omega_{21} \\
&=2\pi i\oint_{a_k}\omega_k(P)\int_{P_0}\sp{P}\Omega_{21}+2\pi i\oint_{b_k}\Omega_{21},
\end{align*}
as the periodicity by shifts of $a$ cycles means
$\oint_{a_k}dg(P)=0$. Similarly we have that
$$
\oint_{b_k+b_k\sp{-1}}\,h(P)dg(P)=0
$$
and therefore (\ref{pre_K2}) yields the expression (\ref{K2}) .\hfill$\Box$

\section{The General $n$ case}
We next inductively establish the general case of Theorem 2.2. We assume
Theorem \ref{genzeros} holds for $n-1$ and consider
\begin{eqnarray}
\int_{P_{0}}^{P_{1}}\omega +\cdots+\int_{P_{0}}^{P_{g+n-1}}\omega
&=&{\bf z} ,\nonumber \\ \int_{P_{0}}^{P_{1}}\Omega_{21}+\cdots
+\int_{P_{0}}^{P_{g+n-1}}\Omega_{21} &=&Z_{2}\nonumber \\
\vdots\qquad\qquad\qquad\qquad&&\vdots \label{cna}
\\ \int_{P_{0}}^{P_{1}} \Omega_{n-1,1} +\cdots
+\int_{P_{0}}^{P_{g+n-1}}\Omega_{n-1,1} &=&Z_{n-1}\nonumber
\\ \int_{P_{0}}^{P_{1}}\Omega_{n1}+\cdots
+\int_{P_{0}}^{P_{g+n-1}}\Omega_{n1} &=&Z_{n} \nonumber ,
\end{eqnarray}
where $P_0\ne \{Q_1,\ldots,Q_n\}$. We wish to show that the
inversion of this map is given by zeros of the generalized theta
function $\varTheta_{n}$ that comes from (\ref{VarTheta}),
(\ref{pars}).

We again begin by supposing that $P_{g+n-1}=P_0$. Then, using the
inductive hypothesis, the solutions of the first $n-2$ of
equations (\ref{cna}) are given in terms of
$\varTheta_{n-1}({\bf z},Z_2,\ldots,Z_{n-1})$, and in this case $Z_n$ is a function of
$\hat{\bf z}=({\bf z},Z_2,\ldots,Z_{n-1})$, namely, the sum of residues,
\begin{align*}
Z_n &= \int_{P_{0}}^{P_{1}}\Omega_{n1}+\cdots
+\int_{P_{0}}^{P_{g+n-2}}\Omega _{n1}\\
&=\sum_{k=1}\sp{g+n-1}\Res_{P=P_k}\left(\int_{P_{0}}^{P}\Omega_{n1}\right)
d\ln\varTheta_{n-1}
(\hat{\bf z}-\mathcal{A}(P))\\
&= \frac{1}{2\pi i}  \oint_{\partial\hat\Gamma_c}
\left(\int_{P_{0}}^{P}\Omega_{n1}\right) d\ln\varTheta_{n-1}
(\hat{\bf z} -\mathcal{A}(P))  .
\end{align*}
Noting that the functions $\varTheta_{n-1} (\hat{\bf
z}-\mathcal{A}(P))$, $\int_{P_{0}}^{P}\Omega_{n1}$ are single
valued on $\hat\Gamma_c$ and taking orientations into account, the
latter expression may be rewritten as
\begin{align*}
Z_n & =\frac{1}{2\pi i}
\left[\oint_{\partial\hat\Gamma}-\int_{\gamma_1+\cdots +\gamma_n}
\right]\left(\int_{P_{0}}^{P}\Omega_{n1}\right)
d\ln\varTheta_{n-1}(\hat{\bf z}-\mathcal{A}(P)) \\
&= \mathcal{K}_n -\frac{1}{2\pi i}\int_{\gamma_1+\cdots+\gamma_n}
\left(\int_{P_{0}}^{P}\Omega_{n1}\right)
d\ln\varTheta_{n-1}(\hat{\bf z}-\mathcal{A}(P)),
\end{align*}
where we set
\begin{equation}
\mathcal{K}_n= \frac{1}{2\pi i}\oint_{\partial\hat\Gamma}
\left(\int_{P_{0}}^{P}\Omega_{n1}\right)
d\ln\varTheta_{n-1}(\hat{\bf z}-\mathcal{A}(P)). \label{defk3}
\end{equation}

According to item 2) of Proposition \ref{propquasi}, for generic
$\hat{\bf z}$ the differential $d\ln\varTheta_{n-1}(\hat{\bf
z}-\mathcal{A}(P))$ has a simple pole (of residue $-1$) at
$Q_2,\dots,Q_{n-2}$ and is regular at $Q_{1,n}$ while
$\Omega_{n1}$ has simple poles at $Q_{1,n}$. Upon an integration
by parts we get
 \begin{align*}
Z_n &= \mathcal{K}_n + \int_{P_0}\sp{Q_2}\Omega_{n1}(P)+
\cdots+\int_{P_0}\sp{Q_{n-1}}\Omega_{n1}(P) +\frac{1}{2\pi
i}\oint_{\gamma_1+\gamma_n} \Omega_{n1}(P)\,\ln \varTheta_{n-1}(\hat{\bf z}-\mathcal{A}(P)) \\
&=\mathcal{K}_n + \int_{P_0}\sp{Q_2}\Omega_{n1}(P)+
\cdots+\int_{P_0}\sp{Q_{n-1}}\Omega_{n1}(P) + \ln \frac{
\varTheta_{n-1}(\hat {\bf z}-\mathcal{A}(Q_n)) } {\varTheta_{n-1}
(\hat{\bf z} -\mathcal{A}(Q_1))} .
\end{align*}
Next, in view of the property (\ref{cnresc}), the above relation yields
\begin{equation*}
 \exp\left(Z_n-\mathcal{K}_n-\sum_{k=2}\sp{n-1}\int_{P_0}\sp{Q_k}\Omega_{n1}(P)\right)
    \theta \left( {\bf z}-K-\sum_{i=1}\sp{n-1}A(Q_i)\right )  +
    \varTheta_{n-1}
(\hat{\bf z}-\mathcal{A}(Q_n))=0 ,
\end{equation*}
which, in view of the recursive definition (\ref{pars}), is
equivalent to $\varTheta_{n}({\bf z},Z_2,\ldots,Z_n)=0$.

Now consider in (\ref{cna}) the generic positive divisor
$P_1,\dots,P_{g+n-1}$ with $P_j\ne Q_1,\ldots,Q_n$. Then upon
replacing $\hat{\bf z}$ by $\hat{\bf z}- \mathcal{A}(P_j) $ (for
any $j\in\{1,\dots,g+n-1\}$) we reduce to the case just considered
and have that $\varTheta_{n}(\hat{\bf z}- \mathcal{A}(P_j) )=0$.
On the other hand, in view of Proposition \ref{propquasi}, the
function $\varTheta_{n} ( {\bf \hat z}- \mathcal{ A} (P))$ has
precisely $g+n-1$ zeros, which therefore coincide with $P_1,\dots,
P_{g+n-1}$. This completes the proof of Theorem \ref{genzeros}.

It remains to derive the expression (\ref{defki}) for
$\mathcal{K}_i$.  Our induction will have established this once we
show that ${\mathcal K}_n$ has this form. Rewriting (\ref{defk3})
we have that
\begin{align*}\mathcal{K}_n&=\frac{1}{2\pi
i}\sum_{k=1}\sp{g}\oint_{a_k+a_k\sp{-1}+b_k+b_k\sp{-1}}\,h(P)\, dg(P)
\end{align*}
where $h(P)\equiv \int_{P_0}\sp{P}\Omega_{n1}$ and $g(P)\equiv
\ln\varTheta_{n-1}(\hat{\bf z}-\mathcal{A}(P))$ are single valued
on $\hat\Gamma_c$. Again letting $h\sp{-}$ denote the value of $h$
on the cycles $a_k\sp{-1}$ or $b_k\sp{-1}$ (and similarly for $g$)
we have that if $P$ lies on $a_k$ then
$$h\sp-(P)=h(P)+\oint_{b_k}\Omega_{n1},\qquad g\sp-(P)=g(P)-2\pi i\,\omega_k(P)$$
while if $b_k$ then $h\sp-(P)=h(P)$, $ g\sp-(P)=g(P)$. Exactly
paralleling the $n=2$ calculation we have that
\begin{align*}
\oint_{a_k+a_k\sp{-1}}\,h(P)dg(P) &= 2\pi
i\oint_{a_k}\omega_k(P)\int_{P_0}\sp{P}\Omega_{n1}+2\pi
i\oint_{b_k}\Omega_{n1},
\end{align*}
and
$$
\oint_{b_k+b_k\sp{-1}}\,h(P)dg(P)=0.
$$
Therefore
$$\mathcal{K}_n=\sum_{k=1}\sp{g}\oint_{a_k}\omega_k(P)\int_{P_0}\sp{P}\Omega_{n1}+
\oint_{b_k}\Omega_{n1},$$ as required.

\begin{proof}[Proof of Theorem \ref{gendiv}]Having the proof of Theorem \ref{genzeros},
we are able to prove Theorem \ref{gendiv}. It assumes that
$f(P)=\varTheta_n ( {\bf \hat z} - \mathcal{A} (P))$ does not
vanish identically. Hence, $f(P)$ has precisely $g+n-1$ zeros
$P_1+\cdots+ P_{g+n-1}$. On the other hand, since ${\bf \hat z}
\in \{ \varTheta_n ( {\bf \hat z} )=0\}$, we have that $f(P_0)=0$,
and one of the above zeros, say $P_{g+n-1}$, coincides with $P_0$.
Then, by Theorem \ref{genzeros}, there is the unique
representation
$$
{\bf \hat z} = {\mathcal A} (P_1) + \cdots + {\mathcal A}
(P_{g+n-2}) + \mathcal{A} (P_0),
$$
which is equivalent to (\ref{theta0}). 
\end{proof}

\paragraph{Remark 2.} Both the theta divisor $\Theta$ and the
set $\{e\}\subset\Theta\subset\Jac(\Gamma)$ for which the function
$f(P)=\theta ( e - {A} (P))$ vanishes identically are interesting
and much studied objects  \cite[Ch. VI \S3]{fk80}. Analogous to
these we have the generalized (translated) theta divisor
$$
\Xi=\{ \varTheta_n ( {\bf \hat z} )=0\}\subset \Jac(\Gamma; Q_1;
Q_2,\dots,Q_n )
$$
and the set $\mathcal{ B}\equiv\{e\} \subset \Jac(\Gamma; Q_1;
Q_2,\dots,Q_n )$ for which $f(P)=\varTheta ( e - \mathcal{ A} (P)
)$ \emph{does} vanish identically. Again $\mathcal{B}\subset \Xi$,
for if $e$ does not belong to $\Xi$ then $f(P)$ cannot vanish
identically. Although our inversion theorems have focused above on
the subset $\{{\bf \hat z}\}\subset \Jac(\Gamma; Q_1;
Q_2,\dots,Q_n )$ for which $f(P)=\varTheta_n ( {\bf \hat z} -
\mathcal{A} (P))$ does not vanish identically, both $\Xi$ and
$\mathcal B$  nonetheless appear interesting objects of study.

\end{document}